\documentclass[jkps,preprint,fleqn,showkeys]{revtex4}
\usepackage{graphicx}
\usepackage{amsmath}
\usepackage{amsthm}
\usepackage{amssymb}
\usepackage{latexsym}
\usepackage{array}
\usepackage{float}
\usepackage{amsfonts}
\usepackage{mathrsfs}
\usepackage{verbatim}
\usepackage{color}

\begin{document}
\setcounter{page}{0}

\title[]{Average fidelity and fidelity deviation in noisy quantum teleportation}

\author{Wooyeong \surname{Song}}\thanks{Wooyeong Song and Junghee Ryu contributed equally to this work}
\affiliation{Department of Physics, Hanyang University, Seoul 04763, Korea}

\author{Junghee \surname{Ryu}}\thanks{Wooyeong Song and Junghee Ryu contributed equally to this work}
\affiliation{Division of National Supercomputing, Korea Institute of Science and Technology Information, Daejeon 34141, Korea}

\author{Kyunghyun \surname{Baek}}
\author{Jeongho \surname{Bang}}\email{jbang@kias.re.kr}
\affiliation{School of Computational Sciences, Korea Institute for Advanced Study, Seoul 02455, Korea}

\received{\today}

\begin{abstract}
We analyze the average fidelity (say, $F$) and the fidelity deviation (say, $D$) in noisy-channel quantum teleportation. Here, $F$ represents how well teleportation is performed on average and $D$ quantifies whether the teleportation is performed impartially on the given inputs, that is, the condition of universality. Our analysis results prove that the achievable maximum average fidelity ensures zero fidelity deviation, that is, perfect universality. This structural trait of teleportation is distinct from those of other limited-fidelity probabilistic quantum operations, for instance, universal-NOT or quantum cloning. This feature is confirmed again based on a tighter relationship between $F$ and $D$ in the qubit case. We then consider another realistic noise model where $F$ decreases and $D$ increases due to imperfect control. To alleviate such deterioration, we propose a machine-learning-based algorithm. We demonstrate by means of numerical simulations that the proposed algorithm can stabilize the system. Notably, the recovery process consists solely of the maximization of $F$, which reduces the control time, thus leading to a faster cure cycle. 
\end{abstract}


\maketitle

\newcommand{\bra}[1]{\left<#1\right|}
\newcommand{\ket}[1]{\left|#1\right>}
\newcommand{\abs}[1]{\left|#1\right|}
\newcommand{\expt}[1]{\left<#1\right>}
\newcommand{\braket}[2]{\left<{#1}|{#2}\right>}
\newcommand{\ketbra}[2]{\left|{#1}\left>\right<{#2}\right|}
\newcommand{\commt}[2]{\left[{#1},{#2}\right]}

\newcommand{\tr}[1]{\mbox{Tr}{#1}}

\newtheorem{theorem}{Theorem}
\newtheorem{lemma}{Lemma}
\newtheorem{definition}{Definition}
\newtheorem{result}{Result}

\section{Introduction}

Quantum teleportation makes possible the deterministic transmission of unknown quantum states from one location to another~\cite{Bennett93}. It has been acknowledged as a fundamental scheme of state transfer. A shared quantum channel between a sender and a receiver is one of the essential ingredients for quantum teleportation, and quantum entanglement in the channel is necessary to ensure that the fidelity is superior to that of classical communication protocols. Moreover, quantum teleportation provides a useful framework to study quantum nonlocality~\cite{Popescu94,Gisin96,Barrett01,Cavalcanti13} and is one of the basic steps in constructing element gates, for example, single-qubit and CNOT gates, which are used in continuous-variable (CV) quantum computation~\cite{Jeong02,Ralph03}. 

Fidelity $f$ is used to measure the closeness between the input and the teleported states. One of the methods to quantify the performance of teleportation based on $f$ without any dependence on the input states involves averaging $f$ over all the possible inputs. The average fidelity $F$, defined as a uniform average of $f$, has been widely used as a relevant input-independent measure of teleportation performance. It is well-known that $F$ can reach $1$ in quantum teleportation, whereas the maximally attainable $F$ is limited in any classical state-transfer scheme without quantum entanglement~\cite{Bennett93}. However, $F$ does not consider universality, which indicates whether teleportation is performed equally for all of the input states. For example, a non universal teleportation protocol would be successful only for a specific set of inputs. This limitation can hinder the use of a teleportation protocol to implement element gates in CV quantum computation~\cite{Jeong02,Ralph03}. To quantify the universality condition, a fidelity deviation $D$ defined in terms of the standard deviation of $f$ was introduced~\cite{unot:Bang12,Bang2018}.

In this study, we analyze the two measures $F$ and $D$ in the context of noisy qudit---a $d$-level quantum system---teleportation. 
We show that perfect universality, $D=0$, is attainable without any dependence on the quantum channel condition, while the maximum average fidelity $F_\text{max}$ is a function of the degree of entanglement in the quantum channel. We prove that the condition of $F_\text{max}$ is adequate to guarantee perfect universality.
For the case of a qubit, that is, $d=2$, we demonstrate a more general and tighter relationship between $F$ and $D$ that addresses the aforementioned behaviors in a clearer manner. Then, we consider a realistic situation in which operational noises can deteriorate the teleportation performance, as represented by a decrease in $F$ and an increase in $D$. We propose a significant machine-learning-based method to alleviate such deterioration. We numerically demonstrate that the proposed machine-learning-based method is effective within a certain rate of noise occurrence. Note that the recovery of $F$ and $D$ can be implemented solely by using the process of $F \to F_\text{max}$. This is because $F_\text{max}$ itself suffices the condition of zero deviation. This feature allows us to reduce the control and/or learning time, which increases the recovery rate.

\section{Average fidelity and fidelity deviation}

We first review the two measures, namely, average fidelity $F$ and fidelity deviation $D$. Consider a map (or a general quantum operation) ${\cal T}: \ket{\phi} \rightarrow \hat{\rho}_\text{out}$, where $\ket{\phi}$ is an input and $\hat{\rho}_\text{out}$ an output. The fidelity $f$ is defined as~\cite{Jozsa94}
\begin{eqnarray}
f = \tr{\left[ \left(\sqrt{\hat{\rho}_\text{t}}\hat{\rho}_\text{out}\sqrt{\hat{\rho}_\text{t}}\right)^{1/2}\right]},
\label{eq:fidelity}
\end{eqnarray}
where $\hat{\rho}_\text{t}$ is a target state that is assumed to be a pure state $\ket{\phi_t}$. Then, formula~(\ref{eq:fidelity}) can be rewritten as
\begin{eqnarray}
f = \bra{\phi_t}\hat{\rho}_\text{out}\ket{\phi_t}.
\end{eqnarray}

In a well-designed ${\cal T}$, almost all of the input states $\ket{\phi}$ are correctly transferred to their corresponding targets $\ket{\phi_t}$. The average fidelity represents {\em how well} a state transfer is performed. It is defined by averaging $f$ over all possible inputs $\ket{\phi}$ as
\begin{eqnarray}
F=\int{\text{d}\phi}\,f,
\label{eq:avgf}
\end{eqnarray}
where $\text{d} \phi$ is the Haar measure that satisfies $\int \text{d} \phi=1$. 
In general, $f$ varies with respect to the input states for a given channel ${\cal T}$. If $f$ is uniform for a task, the task is said to be universal. To devise a measure of universality, we employ the fidelity deviation $D$ in terms of the fluctuation of $f$~\cite{unot:Bang12,Bang2018}:
\begin{eqnarray}
D = \left(\int{\text{d}\phi}\, f^2 - F^2\right)^\frac{1}{2}.
\label{eq:stdd} 
\end{eqnarray}
One can show perfect universality, that is, $D=0$ when $f=F$. Otherwise, it is strictly positive. Additionally,
\begin{eqnarray}
D^2 \leq\int{\text{d}\phi}\,{f} - F^2=F(1-F)\leq\frac{1}{4},
\label{eq:limit_ug}
\end{eqnarray}
where the last equality holds when $F=1/2$. Thus, $D$ is bounded as $0 \le D \le 1/2$. 
 
Before going further, we note that $F=1$ holds iff $f=1$ for all possible inputs, and therefore, $F=1$ implies perfect universality. However, when the attainable maximum of $F$ is less than one ($F_\text{max} < 1$), perfect universality is not guaranteed at $F_\text{max}$ in general. For example, if $F_\text{max} < 1$, then $0 \leq D \leq \sqrt{F(1-F)}$ [see Eq.~(\ref{eq:limit_ug})]. This is true for almost all probabilistic tasks, for instance, universal-NOT \cite{Buzek99,Buzek00-1} and quantum cloning \cite{Buzek96}. Therefore, it is natural to consider the minimization of $D$ independently of the maximization of $F$.

\section{Noisy quantum teleportation}

\begin{figure}[t]
\includegraphics[width=0.8\textwidth]{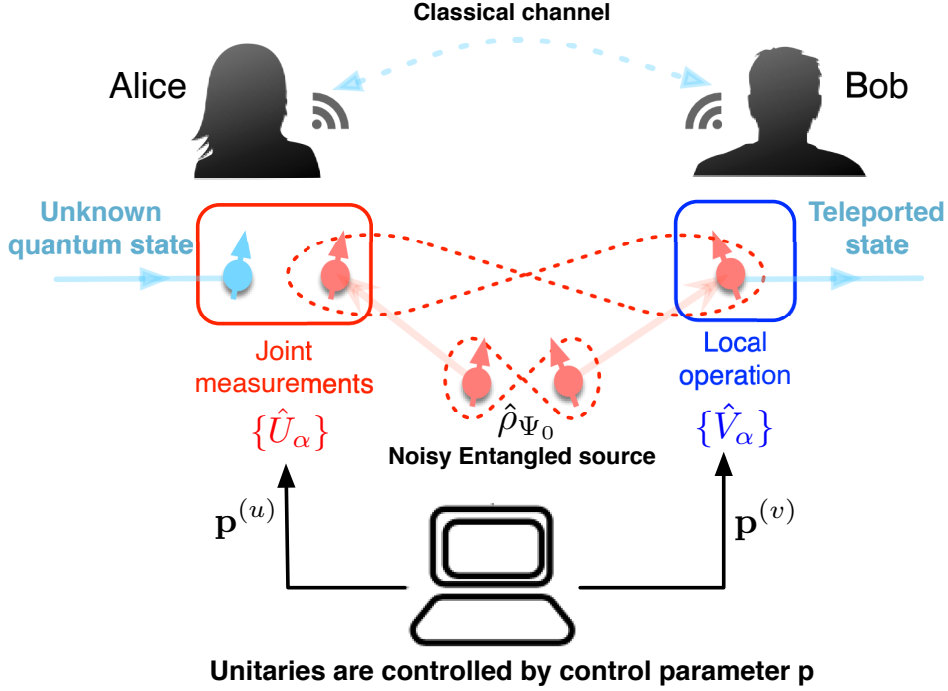}
\caption{Schematic of $d$-dimensional quantum state teleportation. The quantum channel, i.e., shared entangled state $\hat{\rho}_{\Psi_0}$, can be noisy [as in Eq.~(\ref{eq:ws})]. The errors can also arise in $\hat{U}_\alpha$ and $\hat{V}_\alpha$ due to the imperfections of control. It can be cured in our scheme (see Sect.~V for details).}.
\label{fig:tele_scheme}
\end{figure}

In this section, we describe noisy quantum teleportation~\cite{Braunstein00}. 
First, a sender, say Alice, has a $d$-dimensional pure state $\hat{\rho}_\phi = \ketbra{\phi}{\phi}$ that is to be transferred to a receiver, say Bob (in general, Alice has no information about the input state $\hat{\rho}_\phi$).  A maximally entangled state $\ket{\Psi_0}=\frac{1}{\sqrt{d}}\sum_{j=0}^{d-1}\ket{j}\otimes\ket{j}$ is shared by Alice and Bob. Second, Alice performs a joint measurement by using a set of maximally entangled bases $\{\ket{\Psi_\alpha}\}$ ($\alpha=0,1,\ldots,d^2-1$) on the composite system of the unknown state and one of the entangled pairs. The entangled basis is obtained as
\begin{eqnarray}
\ket{\Psi_\alpha} = (\hat{U}_\alpha \otimes \hat{\openone}_d) \ket{\Psi_0}, 
\label{eq:A_m}
\end{eqnarray}
where $\hat{\openone}_d$ is the identity of the $d$-dimensional Hilbert space and $\hat{U}_\alpha$ is a unitary conditioned by completeness, that is, $\sum_{\alpha=0}^{d^2-1}\ket{\Psi_\alpha}\bra{\Psi_\alpha}=\hat{\openone}_{d^2}$~\cite{Braunstein00, Son01}. The outcome of Alice's joint measurement is communicated to Bob through a classical channel. Lastly, Bob applies a local operation $\hat{V}_\alpha$ based on the measurement outcomes received from Alice. Then, the state $\ket{\phi}$ is reproduced on Bob's side (see Fig.~\ref{fig:tele_scheme}). 

Sharing the entanglement between Alice and Bob is the crucial step in ensuring that the better performance is superior to that of any classical protocol~\cite{Horodecki99}. However, under realistic conditions, quantum entanglement often becomes noisy~\cite{Carvalho04,Buric08}, and thus, we need to resolve the noisy case. Herein, we consider the following ($d \times d$)-dimensional noisy entanglement as the quantum channel~\cite{Werner89}:
\begin{eqnarray}
\hat{\rho}_{\Psi_0} = \gamma \ket{\Psi_0}\bra{\Psi_0} + \frac{1-\gamma}{d^2}\hat{\openone}_{d^2},
\label{eq:ws}
\end{eqnarray}
which is a statistical mixture of the maximally entangled state $\ket{\Psi_0}$ and white (symmetric) noise, and $\gamma$ is a fraction of $\ket{\Psi_0}$. 

\section{Analysis of $F$ and $D$ in noisy quantum teleportation}


In this section, we discuss the relationship between the two measures, namely, the average fidelity $F$ and fidelity deviation $D$, according to the noise parameter $\gamma$ and dimension $d$.

\subsection{Qudit-teleportation} 

By definition, the value of $F$ ranges between 0 and 1. However, for a given set of $\gamma$ and $d$, one can restrict the range by using the following relationship:
\begin{result}\label{res:fF}
For a given set of noise parameter $\gamma$ and dimension $d$, $F$ has a relation such that
\begin{eqnarray}
\frac{1}{d} - \frac{\gamma}{d(d+1)} \le F \le  \gamma + \frac{(1-\gamma)}{d}.
\end{eqnarray}
\end{result}
This result is derived from the following extensive analysis of $F$. First, we write the transmitted state $\hat{\rho}_\text{out}$ as
\begin{eqnarray}
\hat{\rho}_\text{out} = \sum_{\alpha=0}^{d^2-1} \left(\hat{\openone}_{d^2} \otimes \hat{V}_\alpha\right) \bra{\Psi^\text{A}_\alpha} \hat{\rho}_\text{tot} \ket{\Psi^\text{A}_\alpha} \left(\hat{\openone}_{d^2} \otimes \hat{V}_\alpha^\dagger\right),
\label{eq:output_init}
\end{eqnarray}
where $\ket{\Psi_\alpha^\text{A}} = \ket{\Psi_\alpha}\otimes\hat{\openone}_d$ denotes Alice's joint measurement, $\hat{\openone}_{d^2} \otimes \hat{V}_\alpha$ represents Bob's local operation, and $\hat{\rho}_\text{tot} = \hat{\rho}_\phi \otimes \hat{\rho}_{\Psi_0}$ is the total initial state of $\hat{\rho}_\phi$ that is to be transferred, where $\hat{\rho}_{\Psi_0}$ is the shared entangled state. We can rewrite Eq.~(\ref{eq:output_init}) as
\begin{eqnarray}
\hat{\rho}_\text{out} = \frac{\gamma}{d^2}\sum_{\alpha=0}^{d^2-1}\hat{X}_\alpha \hat{\rho}_\phi \hat{X}_\alpha^\dagger + \frac{1-\gamma}{d}\hat{\openone}_d,
\end{eqnarray}
where $\hat{X}_\alpha = \hat{V}_\alpha \hat{U}_\alpha^\dagger$. The fidelity $f$ for a given $\hat{\rho}_\phi=\ket{\phi}\bra{\phi}$ is
\begin{eqnarray}
f = \frac{\gamma}{d^2}\sum_{\alpha=0}^{d^2-1}\xi_\alpha + \frac{1-\gamma}{d},
\label{eq:f_d}
\end{eqnarray}
where $\xi_\alpha$ is the fidelity between $\hat{\rho}_\phi$ and $\hat{X}_\alpha\hat{\rho}_\phi\hat{X}_\alpha^\dagger$, and it is expressed as
\begin{eqnarray}
\xi_\alpha = \tr{(\hat{X}_\alpha\hat{\rho}_\phi\hat{X}_\alpha^\dagger \hat{\rho}_\phi}) = \abs{\bra{\phi}\hat{X}_\alpha\ket{\phi}}^2.
\label{eq:df_xi}
\end{eqnarray}

By using the above descriptions, we evaluate $F$ and $D$. First, we consider the average fidelity $F$. From Eqs.~(\ref{eq:avgf}) and (\ref{eq:f_d}), we have
\begin{eqnarray}
F =\int \text{d}\phi \, f = \frac{\gamma}{d^2}\sum_{\alpha=0}^{d^2-1}\int \text{d} \phi \, \xi_\alpha + \frac{1-\gamma}{d}.
\label{eq:F_d}
\end{eqnarray}
The integral $\int \text{d} \phi \, \xi_\alpha$ can be calculated by using a lemma of identity, called Schur's lemma~\cite{Albeverio02,Braunstein00}:
\begin{eqnarray}
\int_G \text{d}g \, \left( \hat{U}_g^\dagger \otimes \hat{U}_g^\dagger \right) \hat{X} \left( \hat{U}_g \otimes \hat{U}_g \right) = a \hat{\openone}_{d^2} + b \hat{P},
\label{eq:sch1}
\end{eqnarray}
where 
\begin{eqnarray}
a = \frac{d^2 \tr{(\hat{X})} - d \tr{(\hat{X}\hat{P})}}{d^2 (d^2 -1)}, \nonumber \\
b = \frac{d^2 \tr{(\hat{X}\hat{P})} - d \tr{(\hat{X})}}{d^2 (d^2 -1)},
\end{eqnarray}
for any operator $\hat{X}$ in the $d \times d$-dimensional Hilbert space. Here, $\text{d}g$ denotes the (normalized) Haar measure on the unitary group $G=U(d)$, satisfying $\int_G \text{d}g = 1$; $\hat{U}_g$ is an irreducible representation of $g \in G$; and $\hat{P}$ denotes the swap operator defined by $\hat{P}\ket{i j} = \ket{j i}$. By applying Schur's lemma, we can compute the integral $\int \text{d}\phi \, \xi_\alpha$ such that
\begin{eqnarray}
\int \text{d}\phi \, \xi_\alpha &=& \bra{\mathbf{0}\mathbf{0}} \int \text{d} \phi \, \left( \hat{U}_\phi^\dagger \otimes \hat{U}_\phi^\dagger \right) \left(\hat{X}_\alpha \otimes \hat{X}_\alpha^\dagger \right) \left( \hat{U}_\phi \otimes \hat{U}_\phi \right) \ket{\mathbf{0}\mathbf{0}} \nonumber \\
    &=& \frac{1}{d(d+1)} \left(\abs{\tr{(\hat{X}_\alpha)}}^2 + d\right),
\label{eq:re_int1}
\end{eqnarray}
where $\ket{\phi}=\hat{U}_\phi\ket{\mathbf{0}}$. Then, we arrive at the final form of $F$ as
\begin{eqnarray}
F = F_\text{max} - \frac{\gamma}{d+1} \left( d - \frac{1}{d^3} \sum_{\alpha=0}^{d^2-1} \abs{\tr{(\hat{X}_\alpha)}}^2 \right),
\label{eq:fF_d}
\end{eqnarray}
where $F_\text{max} = \gamma + \frac{(1-\gamma)}{d}$, which is the maximum value of $F$ for a given $\gamma$. 

We rewrite Eq.~(\ref{eq:fF_d}) in a more useful form as follows:
\begin{eqnarray}
F = \frac{d E + 1}{d+1},
\label{eq:FE}
\end{eqnarray}
where $E$ is defined as
\begin{eqnarray}
E = \frac{1}{d^2}\sum_{\alpha=0}^{d^2-1}\bra{\Psi_\alpha}\hat{\rho}_{\Psi_0}\ket{\Psi_\alpha} = \frac{\gamma}{d^4}\sum_{\alpha=0}^{d^2-1}\abs{\tr{(\hat{X}_\alpha)}}^2 + \frac{1-\gamma}{d^2}.
\label{eq:ef}
\end{eqnarray}
We find that $E$ is maximized when $\tr{(\hat{X}_\alpha)}=d$ (or equivalently, $\hat{X}_\alpha=\hat{\openone}_d$) for all $\alpha$ and minimized when $\tr{(\hat{X}_\alpha)}=0$ for all $\alpha$. Then, the value of $E$ is bounded by $\gamma$ such that
\begin{eqnarray}
\frac{(1-\gamma)}{d^2} \le E \le \gamma + \frac{(1-\gamma)}{d^2},
\end{eqnarray}
where the upper bound $\gamma + \frac{(1-\gamma)}{d^2}$ is called the (fullest) entanglement fraction of the channel~\cite{Albeverio02}. In this manner, by using Eq.~(\ref{eq:FE}) we can finally prove Result~\ref{res:fF}. Notably, this result is consistent with the results obtained in previous studies~\cite{Horodecki99,Albeverio02}.

Next, we consider the fidelity deviation $D$. Moreover, because the range of $F$ is limited, we derive a condition of $D$  for a given $\gamma$ and $d$ as follows.
 \begin{result}
For a given set of noise parameter $\gamma$ and dimension $d$, the fidelity deviation $D$ has a condition such that
\begin{eqnarray}
0 \le D \le \gamma \,\overline{\Delta},
\end{eqnarray}
where $\overline{\Delta} = \frac{1}{d^2}\sum_{\alpha=0}^{d^2-1}\Delta_\alpha$, and 
\begin{eqnarray}
\Delta_\alpha^2 = \int \normalfont \text{d} \phi \, \xi_\alpha^2 - \left(\int \text{d} \phi \, \xi_\alpha\right)^2,
\label{eq:delta_xi}
\end{eqnarray} 
which can be regarded as the fidelity deviation of $\xi_\alpha$.
\end{result}
To prove this result, we express $D$ as follows by using Eqs.~(\ref{eq:f_d}) and (\ref{eq:F_d})
\begin{eqnarray}
D = \left( \int \text{d}\phi \, f^2 - F^2 \right)^\frac{1}{2} = \frac{\gamma}{d^2}\left(\sum_{\alpha,\beta=0}^{d^2-1} C_{\alpha\beta}\right)^{\frac{1}{2}},
\label{eq:D_d}
\end{eqnarray}
where $C_{\alpha\beta}$ are elements of covariance matrix $\mathbf{C}$ given by 
\begin{eqnarray}
C_{\alpha\beta} = \int \text{d}\phi \, \xi_\alpha \xi_\beta  - \int \text{d}\phi \, \xi_\alpha \int \text{d}\phi \, \xi_\beta.
\end{eqnarray}
Note that $\mathbf{C}$ is symmetric, that is, $C_{\alpha\beta} = C_{\beta\alpha}$, and its diagonal elements $C_{\alpha\alpha}$ are equal to $\Delta_\alpha^2$ in Eq. \eqref{eq:delta_xi}. Furthermore, each element of $\mathbf{C}$ is bounded as 
\begin{eqnarray}
\abs{C_{\alpha\beta}} \le \sqrt{\Delta_\alpha^2 \Delta_\beta^2},
\label{eq:ineq_C}
\end{eqnarray}
which is known as the variance-covariance inequality~\cite{Bhatia00}. Then, by using Eq.~(\ref{eq:ineq_C}), we obtain Result 2. The perfect universality $D=0$ can be achieved when $\xi_\alpha$ is constant for all $\alpha$.

Based on the above results, we discuss the conditions of the sets $\{ \hat{U}_\alpha \}$ and $\{ \hat{V}_\alpha \}$ from the viewpoint of achieving the maximum average fidelity and perfect universality. First, we consider the maximization of $F$. According to Eq.~(\ref{eq:fF_d}), $F_\text{max}$ is achieved when
\begin{eqnarray}
\hat{X}_\alpha = \hat{\openone}_d,~\text{or equivalently,}~\hat{U}_\alpha = \hat{V}_\alpha,~(\forall\alpha).
\label{eq:condi_opt}
\end{eqnarray}
That implies that the maximization of $E$ straightforwardly leads to $F_\text{max}$. Therefore, our main result is as follows:
\begin{result}
The condition of Eq.~(\ref{eq:condi_opt}) naturally suffices perfect universality, that is,
\label{result:U}
\begin{eqnarray}
\hat{X}_\alpha = \hat{\openone}_d ~\rightarrow ~\text{the perfect universality}~D=0.
\end{eqnarray}
\end{result}
The proof is simple. By using Eq.~(\ref{eq:df_xi}), we rewrite condition~(\ref{eq:condi_opt}) as
\begin{eqnarray}
\xi_\alpha = 1~(\forall\alpha).
\end{eqnarray}
Then, $\mathbf{C}$ becomes a zero (or null) matrix because $C_{\alpha\beta}=0$ for $\xi_\alpha=const$ ($\forall\alpha$), thus leading to $D=0$, see Eq.~(\ref{eq:D_d}). 
However, the opposite is not always true, that is, perfect universality does not guarantee the maximum fidelity. We emphasize that Result~\ref{result:U} is a non-common trait in limited maximum fidelity tasks, for instance, universal-NOT or cloning~\cite{unot:Bang12}.

\subsection{Qubit-teleportation} 

In the case of a qubit, that is, $d=2$, we can derive a tighter relationship between $F$ and $D$. To this end, we first write the input state $\hat{\rho}_\phi$ in terms of the Bloch representation as
\begin{eqnarray}
\hat{\rho}_\phi  = \frac{1}{2}\left(\hat{\openone}_2 + \boldsymbol{\phi}^\text{T} \boldsymbol{\sigma}\right),
\label{eq:Bphi_st}
\end{eqnarray}
where $\boldsymbol{\phi} = (\phi_x, \phi_y, \phi_z)^\text{T}$ is a Bloch vector of unit norm (i.e., $|\boldsymbol{\phi}|=1$) in three-dimensional real vector space $\mathbb{R}^3$, and $\boldsymbol{\sigma} = (\hat{\sigma}_x, \hat{\sigma}_y, \hat{\sigma}_z)^\text{T}$ is a vector operator whose components $\hat{\sigma}_j$ ($j=x,y,z$) are Pauli operators. Then, we can express $\xi_\alpha$ in Eq.~(\ref{eq:df_xi}) as
\begin{eqnarray}
\xi_\alpha = \frac{1}{2} \left( 1 + \boldsymbol{\phi}^\text{T}\mathbf{R}_\alpha\boldsymbol{\phi} \right),
\label{eq:B_xi}
\end{eqnarray}
where $\mathbf{R}_\alpha$ is a $3 \times 3$ rotation matrix in $\mathbb{R}^3$, whose elements $[\boldsymbol{R}_\alpha]_{jk}$ are given as 
\begin{eqnarray}
[\boldsymbol{R}_\alpha]_{jk} =\frac{1}{2}\tr{(\hat{X}_\alpha \hat{\sigma}_j \hat{X}_\alpha^\dagger \hat{\sigma}_k)}~\text{for}~j,k = x,y,z. 
\end{eqnarray}
The rotation angles $\vartheta_\alpha$ and axes $\mathbf{n}_\alpha$ of $\mathbf{R}_\alpha$ are found in the general expression of a single-qubit unitary as 
\begin{eqnarray}
\hat{X}_\alpha = e^{-i\frac{\vartheta_\alpha}{2}\mathbf{n}_\alpha^\text{T} \boldsymbol{\sigma}} = \cos{\frac{\vartheta_\alpha}{2}}\,\hat{\openone}_2 - i \sin{\frac{\vartheta_\alpha}{2}}\mathbf{n}_\alpha^\text{T} \boldsymbol{\sigma}.
\end{eqnarray}
Then, by using Eqs.~(\ref{eq:F_d}) and (\ref{eq:B_xi}), the $F$ corresponding to $d=2$ can be expressed as
\begin{eqnarray}
F = \frac{\gamma}{4}\sum_{\alpha=0}^3 \frac{1}{2}\left[1 + \int d\boldsymbol{\phi} \left( \boldsymbol{\phi}^\text{T}\mathbf{R}_\alpha\boldsymbol{\phi} \right) \right]  + \frac{1-\gamma}{2},
\label{eq:rwF_1q}
\end{eqnarray}
where $\text{d}\boldsymbol{\phi}$ is the Haar measure over the surface of the Bloch sphere, and it is normalized as $\int \text{d}\boldsymbol{\phi}=1$. Here, we again employ Schur's lemma to calculate the integral $\int \text{d}\boldsymbol{\phi}(\boldsymbol{\phi}^\text{T}\mathbf{R}_\alpha\boldsymbol{\phi})$:
\begin{eqnarray}
\int_G \text{d}g \, \mathbf{O}_g \mathbf{X} \mathbf{O}_g^\text{T} = \frac{1}{r}\tr{(\mathbf{X})}\,\mathbf{I}_{r},
\label{eq:Bschur}
\end{eqnarray}
where $\mathbf{I}_r$ is the identity matrix in $\mathbb{R}^r$, $\mathbf{O}_g$ is an irreducible orthogonal representation of an element $g \in G$, and $\text{d}g$ is the measure, normalized as $\int_g \text{d}g =1$. By applying this lemma to $O(3)$ of three-dimensional rotations, we can calculate the integral in Eq.~(\ref{eq:rwF_1q}) as 
\begin{eqnarray}
\int \text{d}\boldsymbol{\phi} (\boldsymbol{\phi}^\text{T}\mathbf{R}_\alpha\boldsymbol{\phi}) = \frac{1}{3}\tr{(\mathbf{R}_\alpha)},
\label{eq:intB_xi}
\end{eqnarray}
and we immediately obtain the following form of $F$:
\begin{eqnarray}
F = \frac{1}{2} + \frac{\gamma}{24}\sum_{\alpha=0}^{3} \tr{(\mathbf{R}_\alpha)}.
\label{eq:fF_d2}
\end{eqnarray}

Next, we consider $D$. First, we express $D$ as
\begin{eqnarray}
D = \frac{\gamma}{4}\left( \sum_{\alpha=0} ^{3} \Delta_\alpha^2 + \sum_{\alpha \neq \beta} C_{\alpha\beta} \right)^\frac{1}{2}.
\label{eq:rwD_1q}
\end{eqnarray}
Subsequently, we (re)calculate the inequality of Eq.~(\ref{eq:ineq_C}) for $d=2$, and a tighter lower bound can be found as follows (for more details, see appendix~B in Ref.~\cite{unot:Bang12}):
\begin{eqnarray}
-\frac{1}{2}\Delta_\alpha \Delta_\beta \le C_{\alpha\beta}\le \Delta_\alpha \Delta_\beta ~(\alpha \neq \beta),
\label{eq:ineq_Cd2}
\end{eqnarray}
where we obtain the lower bound when the two rotation axes $\mathbf{n}_\alpha$ and $\mathbf{n}_\beta$ are orthogonal to each other, that is, $\mathbf{n}_\alpha^\text{T}\mathbf{n}_\beta = 0$, and the upper bound when $\mathbf{n}_\alpha$ and $\mathbf{n}_\beta$ are parallel or antiparallel, that is, $\mathbf{n}_\alpha^\text{T}\mathbf{n}_\beta = \pm 1$. Then, the upper bound of $D$ is [from Eq.~(\ref{eq:ineq_Cd2})]
\begin{eqnarray}
D \le \frac{\gamma}{4} \sqrt{\sum_{\alpha=0} ^{3} \Delta_\alpha^2 + \sum_{\alpha \neq \beta} \Delta_\alpha \Delta_\beta} = \gamma\overline{\Delta}_{d=2},
\label{eq:rw2D_1q}
\end{eqnarray}
where $\overline{\Delta}_{d=2} = \frac{1}{4}\sum_{\alpha=0}^{3}\Delta_\alpha$. Here, $\Delta_\alpha$ are given in the Bloch form, such that
\begin{eqnarray}
\Delta_\alpha = \frac{1}{2} \left[ \int \text{d} \boldsymbol{\phi}\, \left(\boldsymbol{\phi}^\text{T}\mathbf{R}_\alpha\boldsymbol{\phi}\right)^2  - \frac{1}{9}\tr{(\mathbf{R}_\alpha)}^2 \right]^\frac{1}{2}.
\label{eq:wrB_delta_xi}
\end{eqnarray}
Now, we calculate $\int \text{d} \boldsymbol{\phi}\, (\boldsymbol{\phi}^\text{T}\mathbf{R}_\alpha\boldsymbol{\phi})^2$ by using the product form of Schur's lemma:
\begin{eqnarray}
\int \text{d}g \left(\mathbf{O}_g^\text{T}\otimes\mathbf{O}_g^\text{T}\right)\mathbf{X}\left(\mathbf{O}_g\otimes\mathbf{O}_g\right) = a \mathbf{I}_{d^2} + b \mathbf{D} + c \mathbf{P},
\label{eq:Bschur2}
\end{eqnarray}
where
\begin{eqnarray}
a &=& \frac{(r+1)\tr{(\mathbf{X})} - \tr{(\mathbf{X}\mathbf{D})} - \tr{(\mathbf{X}\mathbf{P})}}{r(r-1)(r+2)}, \nonumber \\
b &=& \frac{-\tr{(\mathbf{X})} + (r+1)\tr{(\mathbf{X}\mathbf{D})} - \tr{(\mathbf{X}\mathbf{P})}}{r(r-1)(r+2)}, \nonumber \\
c &=& \frac{-\tr{(\mathbf{X})} - \tr{(\mathbf{X}\mathbf{D})} + (r+1)\tr{(\mathbf{X}\mathbf{P})}}{r(r-1)(r+2)}. \nonumber
\end{eqnarray}
Here, $\mathbf{P}$ is a swap matrix $\mathbf{P}\,(\mathbf{x}_i \otimes \mathbf{x}_j) = \mathbf{x}_j \otimes \mathbf{x}_i$, or equivalently, $\mathbf{P}=\sum_{i,j=0}^{r-1}\left(\mathbf{x}_j\otimes\mathbf{x}_i \right) \left(\mathbf{x}_i\otimes\mathbf{x}_j \right)^\text{T}$, and $\mathbf{D}=\left(\sum_{i=0}^{r-1} \mathbf{x}_i\otimes\mathbf{x}_i \right) \left( \sum_{j=0}^{r-1}\mathbf{x}_j\otimes\mathbf{x}_j \right)^\text{T}$
where $\{\mathbf{x}_i\}$ is an orthonormal basis set in $\mathbb{R}^r$. Then, by using this lemma, we can rewrite $\int \text{d} \boldsymbol{\phi}\, (\boldsymbol{\phi}^\text{T}\mathbf{R}_\alpha\boldsymbol{\phi})^2$ as
\begin{eqnarray}
\mathbf{x}_{00}^\text{T} \int \text{d} \boldsymbol{\phi} \Big(\mathbf{O}_{\boldsymbol{\phi}}^\text{T} \otimes \mathbf{O}_{\boldsymbol{\phi}}^\text{T} \Big) \left(\mathbf{R}_\alpha \otimes \mathbf{R}_\alpha\right) \Big(\mathbf{O}_{\boldsymbol{\phi}} \otimes \mathbf{O}_{\boldsymbol{\phi}} \Big) \mathbf{x}_{00},
\label{eq:stdD_1q-int}
\end{eqnarray}
where $\mathbf{x}_{00}=\mathbf{x}_0 \otimes \mathbf{x}_0$ and $\boldsymbol{\phi} = \mathbf{O}_{\boldsymbol{\phi}}\mathbf{x}_0$. Eq.~(\ref{eq:wrB_delta_xi}) is then calculated and we have
\begin{eqnarray}
\Delta_\alpha = \frac{1}{2\sqrt{5}}\left(1-\frac{1}{3}\tr{(\mathbf{R}_\alpha)}\right),
\end{eqnarray}
where we used the following properties: 
\begin{eqnarray}
\tr{(\mathbf{R}_\alpha\otimes\mathbf{R}_\alpha)} &=& \tr{(\mathbf{R}_\alpha)}^2, \nonumber \\
\tr{(\mathbf{R}_\alpha\otimes\mathbf{R}_\alpha\,\mathbf{D})} &=& \tr{(\mathbf{R}_\alpha\mathbf{R}_\alpha^\text{T})}=\tr{(\mathbf{I}_3)}=3, \nonumber \\
\tr{(\mathbf{R}_\alpha\otimes\mathbf{R}_\alpha\,\mathbf{P})} &=& \tr{(\mathbf{R}_\alpha^2)}.
\end{eqnarray}
Consequently, we can derive a tighter relationship between $F$ and $D$ as follows:
\begin{eqnarray}
D \le \frac{1}{\sqrt{5}} \left( F_\text{max} - F \right),
\label{eq:upb_d2}
\end{eqnarray}
where $F_\text{max} = \frac{1 + \gamma}{2}$. This confirms Result~\ref{result:U}, that is, $D=0$ iff $F=F_\text{max}$. Here, we can prove that the maximum value of $D$, that is, the worst case of universality, implies the minimum average fidelity, $F_\text{min}=\frac{1}{2} - \frac{\gamma}{6}$.

\section{Machine-learning-based stabilization of control in teleportation}

Result~\ref{result:U} implies the following: [{\bf T.1}] {\em The maximization of $F$ naturally includes the minimization of $D$ in noisy teleportation.} This is indeed a structural trait of teleportation. We conjecture that quantum teleportation degraded by operational noises can be effectively cured and further provide a benefit of utilizing the aforementioned trait [{\bf T.1}] can be achieved. To investigate this, we assume the operational noises in the control of Alice's joint measurements (i.e., $\{\hat{U}_\alpha\}$) and Bob's local operations (i.e., $\{\hat{V}_\alpha\}$), which deteriorate $F$ and $D$. We propose a machine-learning-based algorithm to stabilize the teleportation system against noise. For qubit teleportation (i.e., $d=2$), we demonstrate by means of numerical simulations that our machine-learning-based algorithm can cure the deterioration of both $F$ and $D$ solely by maximizing $F$ (see Fig.~\ref{fig:tele_scheme}).

\subsection{Effects of operational noise} 

In general, a unitary operation in the $d$-dimensional Hilbert space can be parameterized as follows: 
\begin{eqnarray}
\label{eq:unit_op}
\hat{U}(\mathbf{p}) = \exp{(-i\,\mathbf{p}^\text{T}\mathbf{G})},
\end{eqnarray}
where $\mathbf{p}=(p_1,p_2,\ldots,p_{d^2-1})^\text{T}$ is a $(d^2-1)$-dimensional real vector and $\mathbf{G}=(\hat{g}_1, \hat{g}_2, \ldots,\hat{g}_{d^2-1})^\text{T}$ is a vector operator whose components are SU($d$) group generators $\hat{g}_j$ ($j=1,2,\ldots d^2-1$)~\cite{Hioe81,Son04,Bang08}. Here, the components of $\mathbf{p}$ correspond to a set of control parameters in a real experiment, such as multiport beam splitters and phase shifters in linear optical systems~\cite{Reck94} or the pulse sequences of solid system qubits~\cite{Lee00}. The operational noises of $\{\hat{U}_\alpha\}$ and $\{\hat{V}_\alpha\}$ can be formulated as follows: the control parameters of $\hat{U}_\alpha$ and $\hat{V}_\alpha$ fluctuates such that
\begin{eqnarray}
\label{eq:err} 
\mathbf{p} \rightarrow \mathbf{p} +\eta\,\boldsymbol\epsilon,
\end{eqnarray}
where $\boldsymbol\epsilon = (\epsilon_1, \epsilon_2, \ldots, \epsilon_{d^2-1})^\text{T}$ is a vector of the stochastic errors $\epsilon_j \in [-\pi, \pi]$. The factor $\eta \in [0,1]$ is casted to represent the degree of immaturity in control.

\begin{figure}[t]
\includegraphics[angle=270,width=0.37\textwidth]{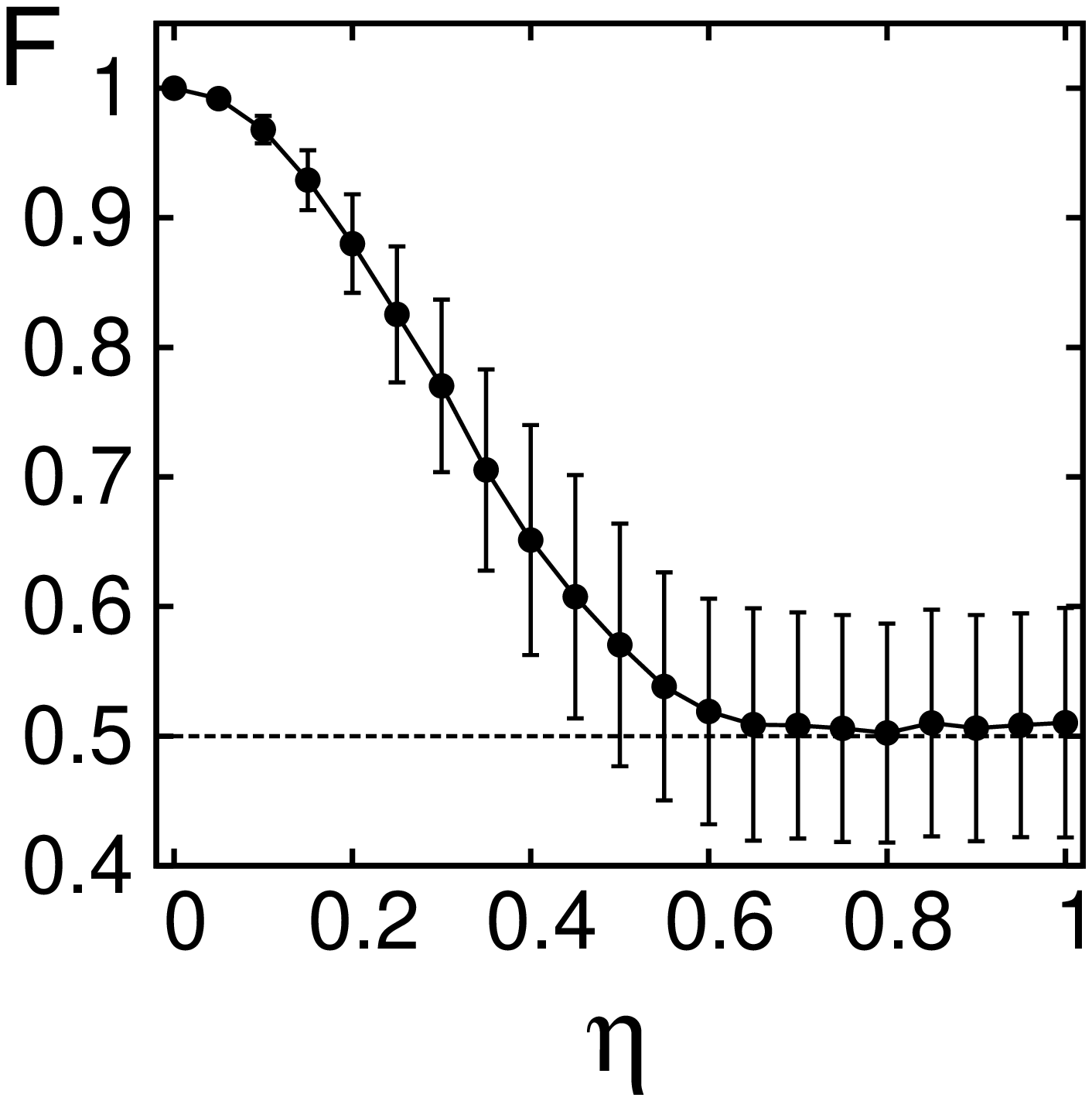}
\includegraphics[angle=270,width=0.37\textwidth]{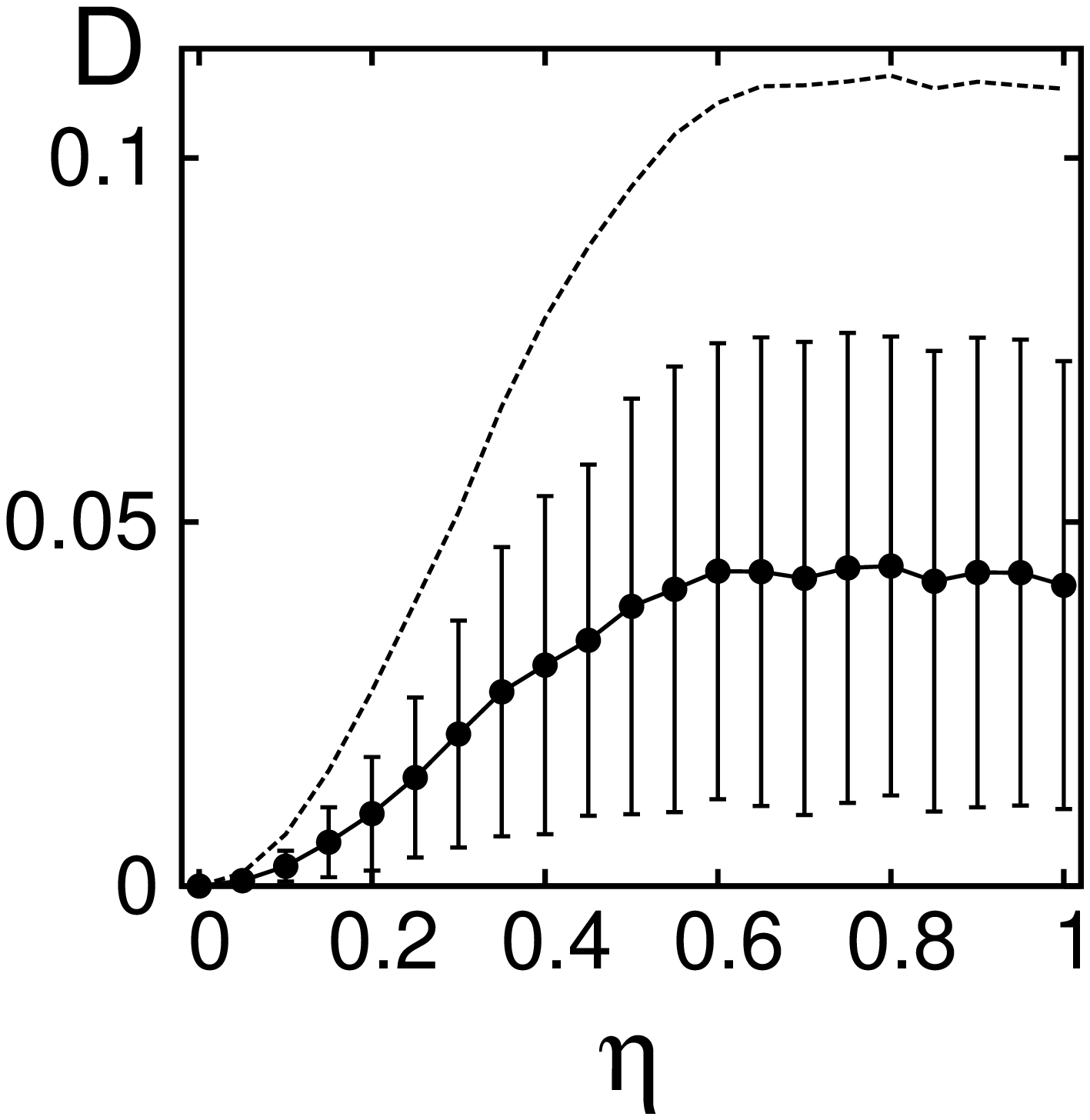}
\caption{Deteriorations of $F$ (left) and $D$ (right) with respect to $\eta$. Each data point is created by averaging $10^4$ simulations. The error bar denotes the standard deviation. The teleportation system breaks down due to operational noise, as indicated by $F=1 \rightarrow F_\text{err} < 1$ and $D=0 \rightarrow D_\text{err} > 0$. The dashed lines are created based on simulations of the random protocol.}
\label{fig:FD_operr}
\end{figure}

We investigate the effects of operational noises on qubit teleportation. First, we assume that the protocol is already set with the optimal condition as in Eq.~(\ref{eq:condi_opt}), and in this case, operational noise occurs continuously in both $\{\hat{U}_\alpha\}$ and $\{\hat{V}_\alpha\}$. To understand the effects of operational noise, we perform Monte-Carlo simulations with increasing $\eta$. The results are shown in Fig.~\ref{fig:FD_operr}, where each data point is created by averaging $10^4$ simulations. $F$ decreases from $1$ to $F_\text{err}$, and $D$ increases from $0$ to $D_\text{err}$. The fullest deterioration of $F$, that is, $F_\text{err} \simeq 0.5102$ is approximately $98\%$ to $F_\text{rand}=\frac{1}{2}$, whereas $D_\text{err}$ increases to $\simeq 0.0431$, or by approximately $39\%$, to $\frac{1}{2}D_\text{err}^\text{max} \simeq \frac{1}{4\sqrt{5}}$ even in the worst case of $F_\text{err} \simeq \frac{1}{2}$. This implies that the deterioration of $F$ is more conspicuous than that of $D$. Here, $F_\text{rand}$ is the average fidelity of the purely random protocol and $\frac{1}{2}D_\text{err}^\text{max}$ represents half (mean) of the maximum fidelity deviation at $F_\text{err}$, that is, $\frac{1}{2}D_\text{err}^\text{max} = \frac{1}{2\sqrt{5}}(F_\text{max}-F_\text{err})$ (see the dashed line in Fig.~\ref{fig:FD_operr}). 

\subsection{Machine-learning-based algorithm for stabilization of control} 

To cure the unstable control, we propose a machine-learning-based algorithm built on the so-called differential evolution concept~\cite{unot:Bang12}, where the control parameters of $\{\hat{U}_\alpha\}$ and $\{\hat{V}_\alpha\}$ are allowed to evolve during the process. The algorithm runs as follows: First, $N_\text{pop}$ sets of the control parameter vectors are prepared as the candidate solutions $\{\mathbf{p}^{(u)}_{\alpha,i}, \mathbf{p}^{(v)}_{\alpha,i}\}$, where $\mathbf{p}^{(u)}_{\alpha,i}$ and $\mathbf{p}^{(v)}_{\alpha,i}$ are respectively the control parameter vectors of the candidate operations $\hat{U}_{\alpha,i}$ and $\hat{V}_{\alpha,i}$ ($i=1,2,\ldots,N_\text{pop}$). Thus, we have $2 d^2 N_\text{pop}$ parameter vectors. Then, the prepared sets of the candidate solutions are allowed to evolve through the following steps: (1) We generate $2N_\text{pop}$ mutant vectors $\boldsymbol\nu^{(u,v)}_{\alpha,i}$ for $\hat{U}_{\alpha,i}$ and $\hat{V}_{\alpha,i}$ according to 
\begin{eqnarray}
\boldsymbol\nu^{(u,v)}_{\alpha,i} = \mathbf{p}^{(u,v)}_{\alpha,a} + W \left(\mathbf{p}^{(u,v)}_{\alpha,b} - \mathbf{p}^{(u,v)}_{\alpha,c}\right),
\end{eqnarray}
where $\mathbf{p}^{(u,v)}_{\alpha,a}$, $\mathbf{p}^{(u,v)}_{\alpha,b}$, and $\mathbf{p}^{(u,v)}_{\alpha,c}$ are randomly selected for $a,b,c \in \{1,2,\ldots,N_\text{pop}\}$. These vectors are selected to be different from each other. The free parameter $W$, also called a differential weight, is a real and constant number. (2) Thereafter, all $2 d^2 N_\text{pop}$ parameter vectors, 
\begin{eqnarray}
\mathbf{p}^{(u)}_{\alpha,i} =
\begin{pmatrix}
p^{(u)}_{1} \\
p^{(u)}_{2} \\
\vdots \\
p^{(u)}_{d^2-1},
\end{pmatrix}_{\alpha,i},~
\mathbf{p}^{(v)}_{\alpha,i} =
\begin{pmatrix}
p^{(v)}_{1} \\
p^{(v)}_{2} \\
\vdots \\
p^{(v)}_{d^2-1},
\end{pmatrix}_{\alpha,i}
\end{eqnarray}
are reformed to trial vectors, 
\begin{eqnarray}
\boldsymbol{\tau}^{(u)}_{\alpha,i} =
\begin{pmatrix}
\tau^{(u)}_{1} \\
\tau^{(u)}_{2} \\
\vdots \\
\tau^{(u)}_{d^2-1},
\end{pmatrix}_{\alpha,i},~
\boldsymbol{\tau}^{(v)}_{\alpha,i} =
\begin{pmatrix}
\tau^{(v)}_{1} \\
\tau^{(v)}_{2} \\
\vdots \\
\tau^{(v)}_{d^2-1},
\end{pmatrix}_{\alpha,i}
\end{eqnarray}
by the rule: For each $j = 1,2,\ldots,d^2-1$,
\begin{eqnarray}
\label{eq:crossover}
\left\{
\begin{array}{ll}
\tau^{(u,v)}_{j} \leftarrow p^{(u,v)}_{j} & ~~\text{if}~R_j > C_r,\\
\tau^{(u,v)}_{j} \leftarrow \nu^{(u,v)}_{j} & ~~\text{otherwise}, \\
\end{array}
\right.
\end{eqnarray}
where $R_j \in [0, 1]$ is a randomly generated number, and the crossover rate $C_r$ is another free parameter ranging between $0$ and $1$. Note that these free parameters $W$ and $C_r$ are set to achieve the best learning efficiency. (3) Finally, the control parameter vectors are evaluated by using the fitness criteria, that is, how well do the given parameters fit to the protocol. More specifically, $\{\boldsymbol\tau^{(u)}_{\alpha,i}, \boldsymbol\tau^{(v)}_{\alpha,i}\}$ are taken if they yield a higher level of fitness; if not, $\{\mathbf{p}^{(u)}_{\alpha,i}, \mathbf{p}^{(v)}_{\alpha,i}\}$ are retained. In our algorithm, we extract the best fitness among $N_\text{pop}$ and retain the corresponding parameters $\{\mathbf{p}_{\alpha,\text{best}}, \mathbf{p}_{\alpha,\text{best}}\}$. Steps (1)-(3) are then repeated.

\begin{figure}[t]
\includegraphics[angle=270,width=0.37\textwidth]{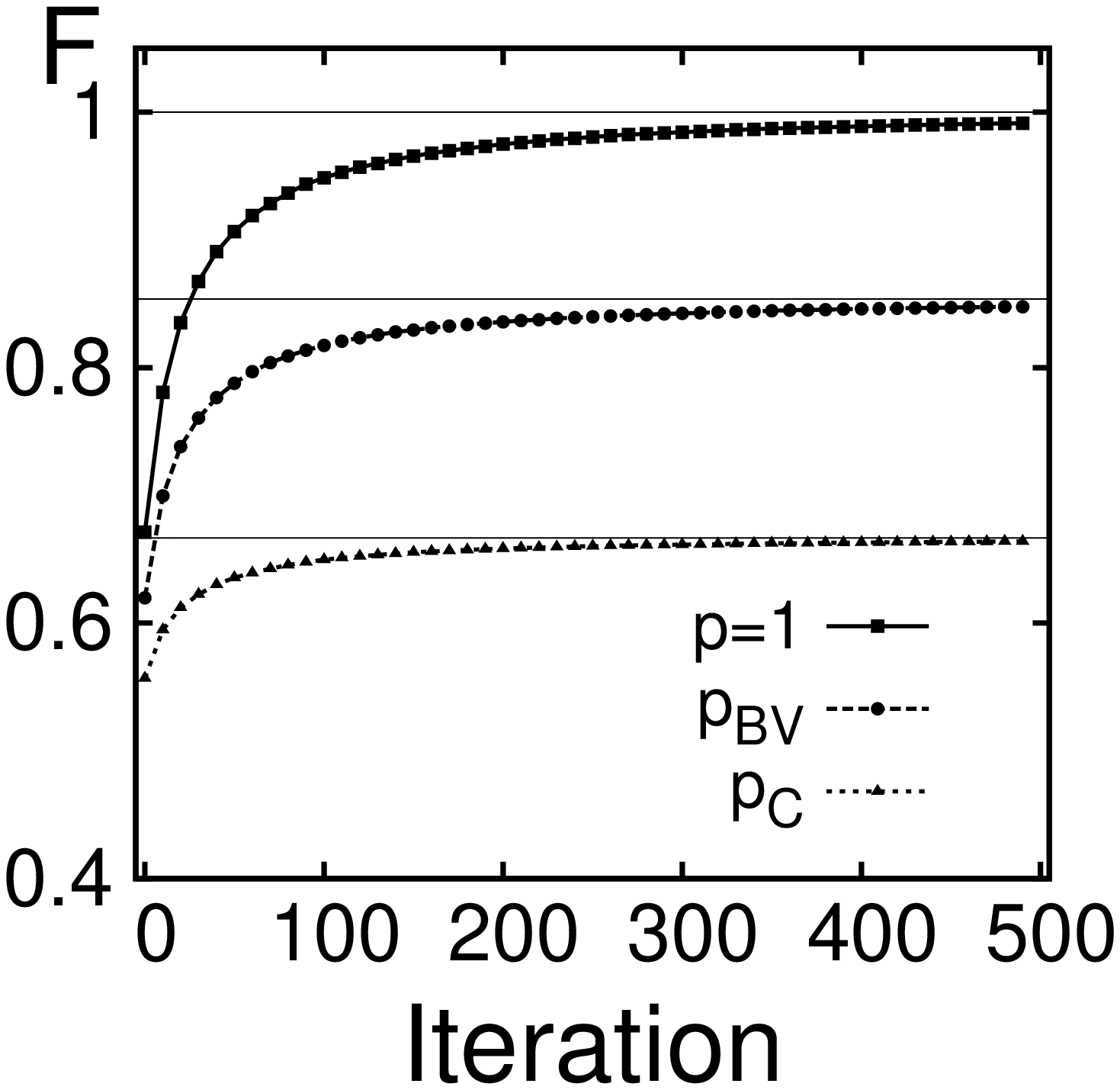}
\includegraphics[angle=270,width=0.37\textwidth]{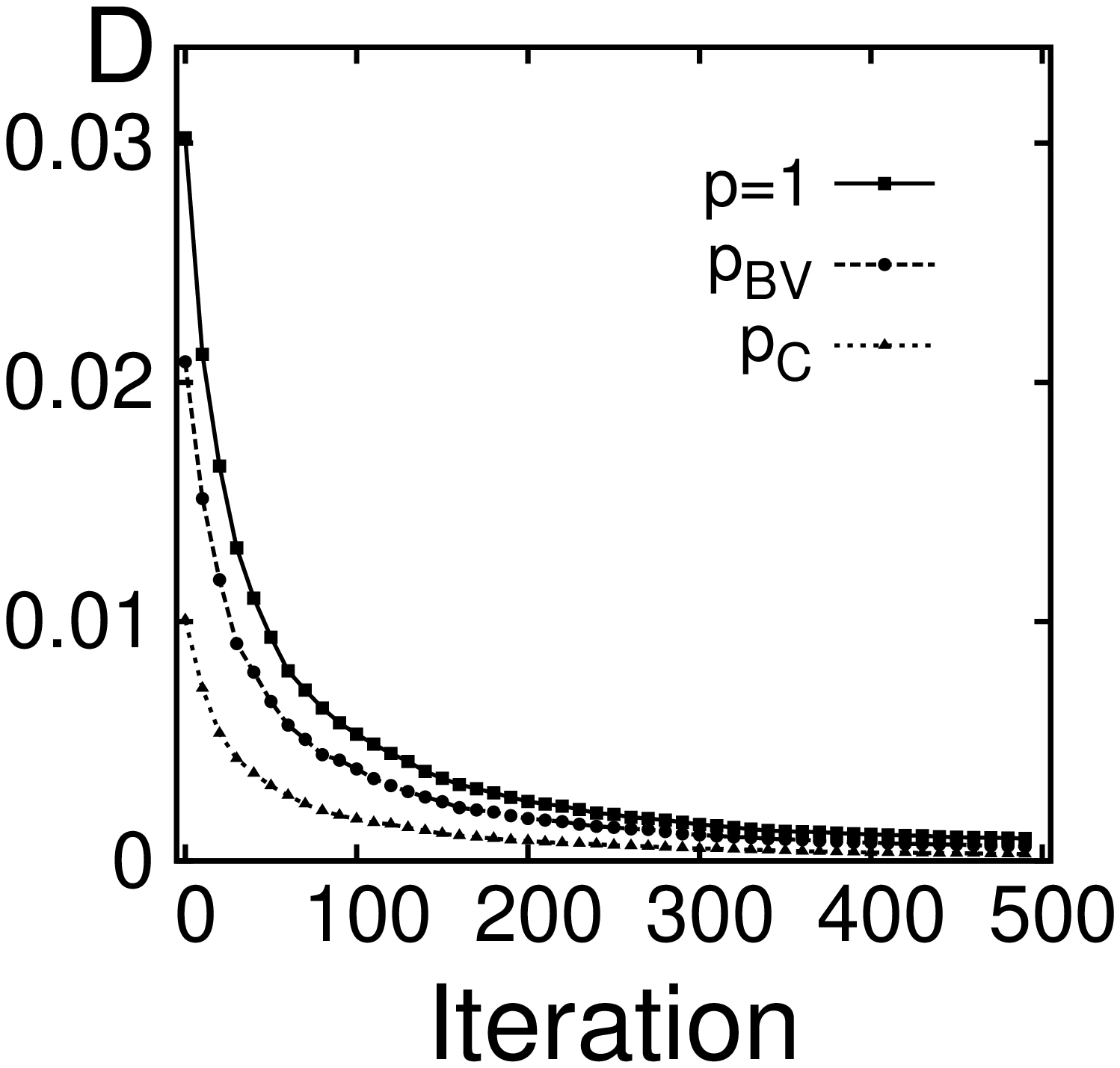}
\caption{Remediation of fully broken teleportation. The graphs of $F$ (left) and $D$ (right) are plotted based on the results of the numerical simulations. Here, we consider three cases: $\gamma=\gamma_\text{C}=\frac{1}{3}$, $\gamma=\gamma_\text{BV}=\frac{1}{\sqrt{2}}$, and $\gamma=1$. Each data point is created based on $10^4$ repeating simulations.}
\label{fig:de_find}
\end{figure}

We investigate numerically whether even a fully broken teleportation system can be cured using the proposed machine-learning-based algorithm. The numerical simulation is performed for $d=2$. Here, we take $N_\text{pop}=100$, and the free parameters of our algorithm are selected such that: $W=0.5$ and $C_r = 0.1$. To use the structural trait [{\bf T.1}] of teleportation, we define fitness in terms of $F$; in other words, there is no minimization of $D$. Note that in general, fitness should be defined as a function of $F$ and $D$ (for example, see Ref.~\cite{unot:Bang12}). Such a setting is indeed beneficial, as described later. In Fig.~\ref{fig:de_find}, we present the results in the form of graphs of $F$ and $D$ for three cases: $\gamma=\gamma_\text{C}=\frac{1}{3}$, $\gamma=\gamma_\text{BV}=\frac{1}{\sqrt{2}}$, and $\gamma=1$. Here, $\gamma_\text{C}$ denotes the condition of the separability of the channel and $\gamma_\text{BV}$ is the critical value that the entanglement of the channel allows the violation of CHSH inequality. Each data point is created based on $10^4$ repeated simulations. The results indicates that the broken teleportation system can be recovered; $F$ approaches $F_\text{max}$, and $D$ decreases to zero. Specifically, we obtain ($F \simeq 0.6640$, $D \simeq 0.0002$) for $\gamma=\gamma_\text{C}$, ($F \simeq 0.8478$, $D \simeq 0.0005$) for $\gamma=\gamma_\text{BV}$, and ($F \simeq 0.9915$, $D \simeq 0.0009$) for $\gamma=1$. 

\begin{figure}[t]
\includegraphics[angle=270,width=0.37\textwidth]{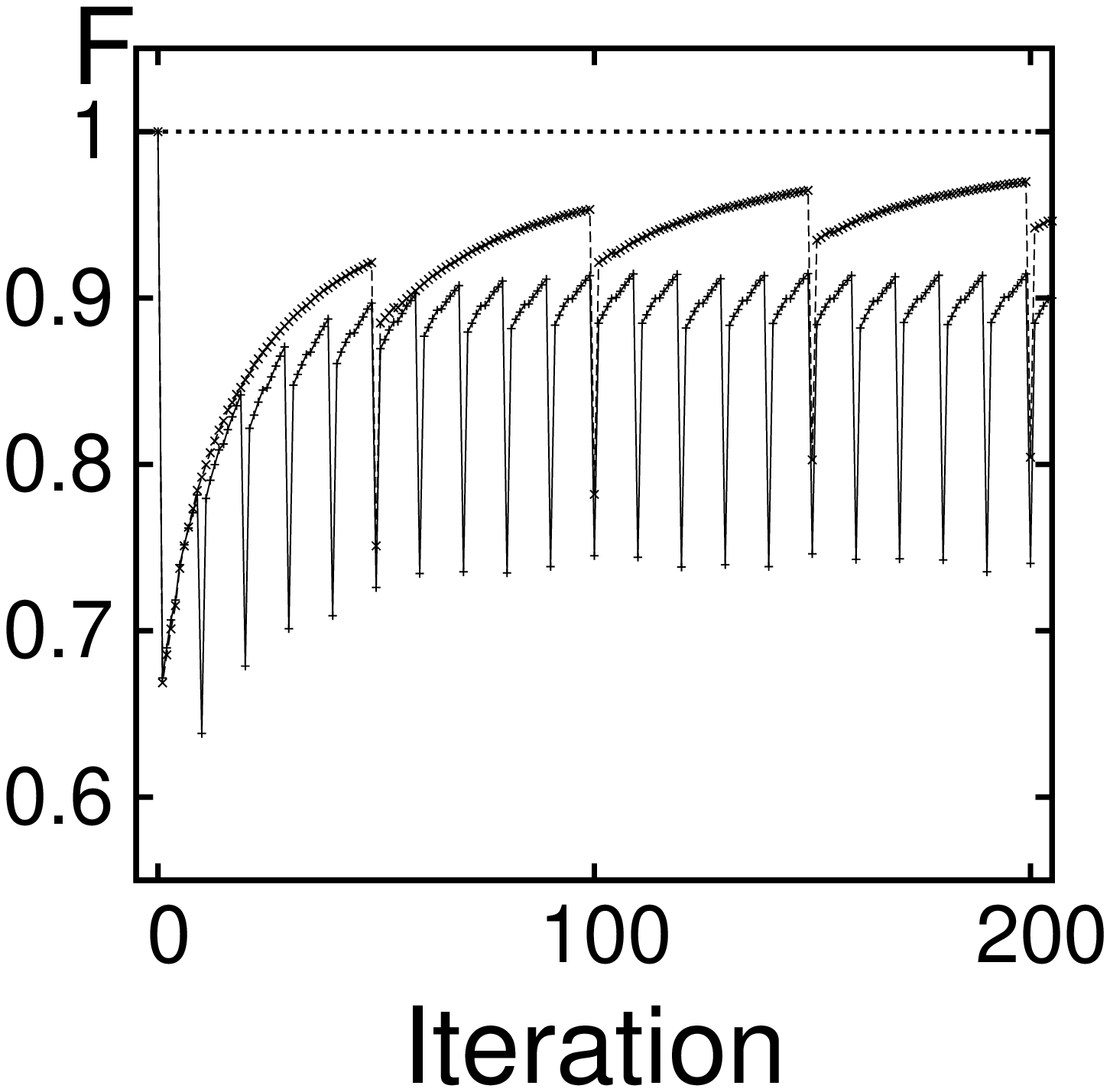}
\includegraphics[angle=270,width=0.37\textwidth]{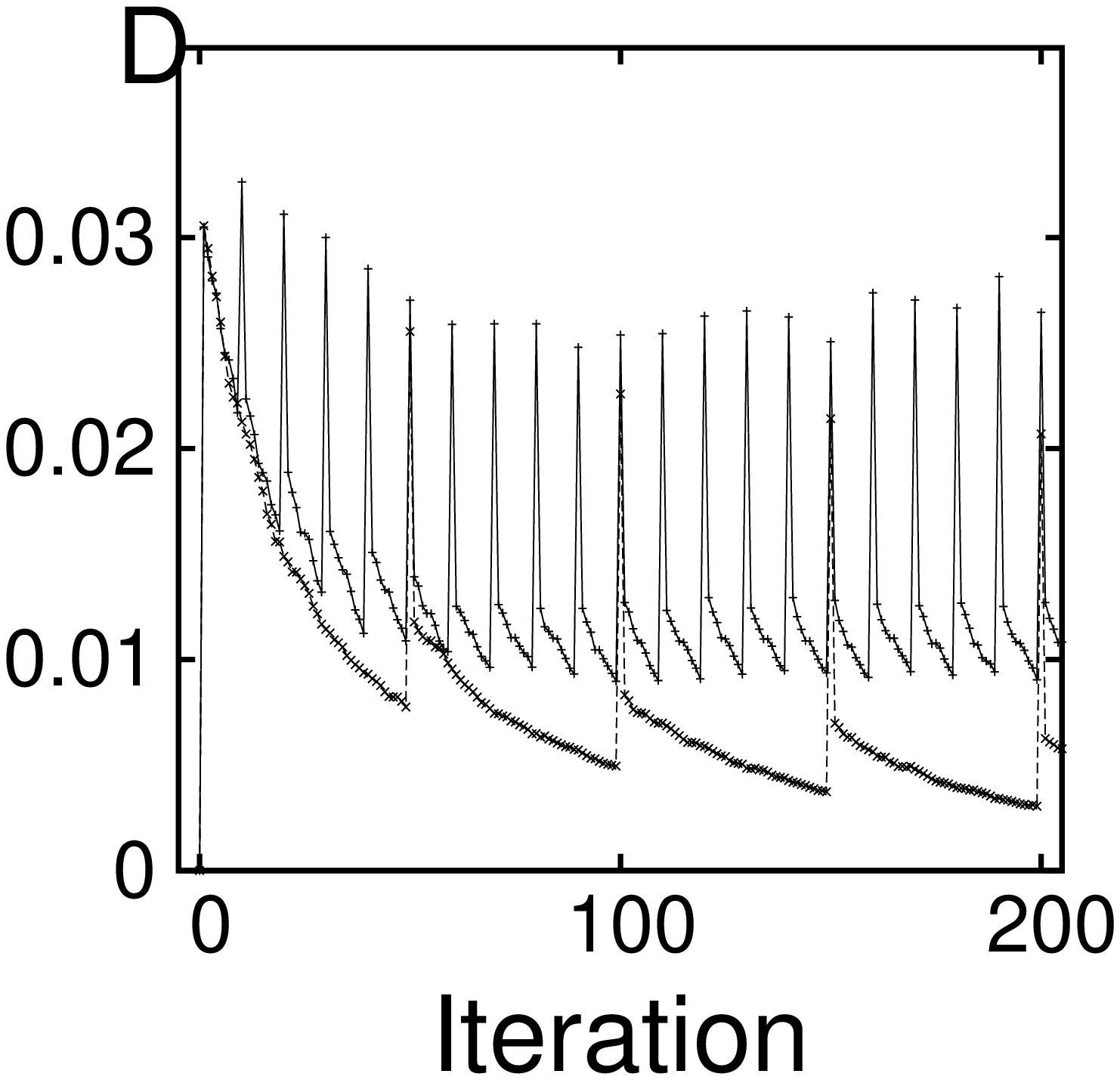}
\caption{Real-time remediation of control fluctuations. The results of the numerical simulations are given as graphs of $F$ (left) and $D$ (right). Each data point is created based on $10^4$ simulations. It is assumed that fluctuation occurs after every $10$ or $50$ iterations of our algorithm. For more convincing analysis, we consider the worst case, namely, $\eta =1$. It is observed that the teleportation can be cured continuously.}
\label{fig:comp_err}
\end{figure}

We further investigate through numerical simulations whether the system can be stabilized, by assuming that the fluctuation occurs abruptly after intervals of some iterations of our algorithm. Note that such a model is realistic~\cite{Viola99, Khaneja01}. Here, we consider the scenarios in which the controls fluctuate after every $10$ and $50$ iterations. For more faithful and confident analysis, we consider the scenario with the worst deterioration, that is, $\eta=1$. Fitness is defined solely by $F$. In Fig.~\ref{fig:comp_err}, we present our simulation results. $F$ and $D$ deteriorate to their fullest extents, but the system is cured continuously. 

\section{Summary and remarks}

We analyzed the average fidelity and fidelity deviation for noisy teleportation. We proved that teleportation can be zero fidelity deviations (or equivalently, the perfect universality) independently of the quantum channel condition, while the achievable maximum average fidelity is limited by the fraction of entanglement in the channel. Based on these analyses, we derived Result~\ref{result:U}: the maximum average fidelity ensures perfect universality in quantum teleportation. For the case of $d=2$, we derived a tighter relationship between the two measures. Taking into account other realistic noises, namely, the fluctuations in system control, we proposed a machine-learning-based algorithm to stabilize teleportation. We demonstrated by means of numerical simulations that even the fullest deteriorations can be cured.  
It is remarkable that the process of fidelity maximization guarantees the minimization of fidelity deviation without additional processes.
The aforementioned trait (coming from Result~\ref{result:U}) is indeed beneficial to reduce the algorithm time and realize faster system remediation; in fact, if the minimization of fidelity deviation was considered in the algorithm, we may not have obtained a cure cycle (i.e., sufficient time for iterations to cure abrupt fluctuations). Such a gain is expected to be more conspicuous in large-$d$ teleportation. 

\section*{Acknowledgments}

JB thanks to M. Wie\'{s}niak and T. V\'{e}rtesi for discussions. WS and JB thank to the financial support of the National Research Foundation (NRF) of Korea Grants (no.~2019R1A2C2005504 and no.~NRF-2019M3E4A1079666), funded by the MSIP (Ministry of Science, ICT and Future Planning), Korea government. JR acknowledges the National Research Foundation of Korea (NRF) Grants no. NRF-2020M3E4A1079792). JB and KB was supported by KIAS Individual Grants (no.~CG061003 and no.~CG074701), respectively. JB and KB was supported by the NRF Grant funded by the Korea government(MSIT) (no.~2020M3E4A1079939).


\end{document}